\documentclass[aps,prb,twocolumn,floatfix,amsmath,superscriptaddress,showpacs]{revtex4}
\usepackage{amsmath}
\usepackage{epsfig,epsf}
\usepackage{subfigure}

\def\be{\begin{equation}}
\def\ee{\end{equation}}
\def\ba{\begin{eqnarray}}
\def\ea{\end{eqnarray}}

\begin{document}

\title
{Superconductivity in Inhomogeneous Hubbard Models}

\author{Wei-Feng Tsai} 
\affiliation{Department of Physics and Astronomy, UCLA, Los Angeles, CA 90095,  
USA} 
\author{Steven~A. Kivelson} 
\affiliation{Department of Physics and Astronomy, UCLA, Los Angeles, CA 90095,  
USA}
\affiliation{Department of Physics, Stanford University, Stanford, CA 94305,  
USA} 

\date{\today}

\begin{abstract}
We present a controlled perturbative approach to the low temperature phase diagram of highly inhomogeneous Hubbard models in the limit of small coupling, $t'$, between clusters.  We apply this to the dimerized and checkerboard models.  The dimerized model is found to behave like a doped semiconductor, with a Fermi-liquid groundstate with parameters ({\it e.g.} the effective mass) which are smooth functions of the Hubbard interaction, $U$.  By contrast, the checkerboard model has a nodeless $d$-wave superconducting state (preformed pair condensate, $d$-BEC) for $0 < U < U_c$, which smoothly crosses over to an intermediate BCS-like superconducting phase ($d$-BCS), also with no nodal quasi-particles, for $|U - U_c| < {\cal O}(t^\prime)$, which gives way to a Fermi liquid phase at large $U > U_c = 4.58$.

\end{abstract}

\pacs{74.20.Mn, 71.10.Fd}

\maketitle 

In this paper, we report a study of inhomogeneous Hubbard models, Eq. (\ref{eq:ihm}), in which the lattice is broken up into a periodic array of weakly coupled clusters. We focus on the case of small ``doping,''  $|x| \ll 1$, where $1\pm x$ is the number of electrons per site, and on the limit in which the coupling between clusters, $t^\prime$, is much less than the relevant energy scales within a cluster. Exploiting these small parameters, we obtain a well controlled solution of the ground-state and low energy excited states.    

There are two purposes of this study:  1) With $t^\prime$ as the small parameter, we can trace the non-perturbative evolution of the electronic structure as a function of the strength of the Hubbard interaction, $U$, all the way from the weak to the strong coupling limit.  2) In light of the increasing evidence \cite{rmp} that some form of self-organized electronic inhomogeneity is widespread in the cuprate superconductors, it is reasonable to explore the circumstances in which high temperature superconductivity from purely repulsive interactions may be enhanced by certain forms of inhomogeneity  \cite{sudip,carlson,Arrigoni,ivar,fradkinreview,arita}. 

We develop a general strategy for such problems which we apply explicitly to the case of the dimerized Hubbard model (Fig. 1(a)) and the checkerboard Hubbard model (Fig. 1(b)). In both  these cases, the undoped ($x=0$) ``parent'' Mott insulating system has a unique, insulating ground-state with a large spin-gap, $\Delta_s$:

{\noindent {\bf1) }}  The doped dimerized Hubbard model has a spectacularly featureless phase diagram. At energies small compared to $\Delta_s$, it behaves like a doped semiconductor, with a small Fermi surface enclosing a Luttinger volume equal to $x$, and with an effective mass which changes by  a factor of 2 as $U$ is increased from $U=0$ to $U \gg 1$. If some form of attractive interaction is added to the dimerized Hubbard model, such as an additional nearest-neighbor exchange energy, $J$, there is a transition to a singlet superconducting phase for sufficiently large $J$, as indicated in Fig. 2. However, the superconducting state has mixed $d$- and $s$-wave symmetry, and a full gap to quasiparticle excitations. 

{\noindent {\bf 2)}}  The doped checkerboard Hubbard model exhibits four distinct zero-temperature phases as a function of $U$, as shown in Fig. 3(a): For $0<U < U_c \approx 4.58$, the system is superconducting while for $U>U_c$ it is a non-superconducting Fermi liquid. The superconducting state has $d$-wave symmetry. Despite this, the quasiparticle spectrum in the superconducting state is fully gapped, even in the range of $U$ within ${\cal O}(t^\prime)$ of $U_c$, where BCS-like $d$-wave superconducting state occurs (explain below).
The Fermi liquid phase is unusual in the sense that there are two degenerate bands (``flavors'') of fermions with plus and minus chirality (in addition to the two spin polarizations). At very large $U > 18.6$ there is an additional transition to a Fermi liquid with spin 3/2 quasiparticles.

\begin{figure}[t]
\includegraphics[width=5.8cm]{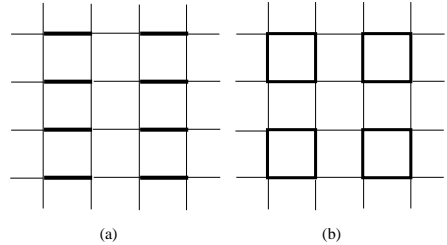}
\caption{{\footnotesize Schematic representation of the 
(a) dimerized and (b) checkerboard models.
The hopping matrix elements are $t=1$ on the bold lines and $t^\prime \ll 1$
on the thin lines.
}}
\end{figure}

It is also worth noting that inhomogeneous systems naturally exhibit precursor superconducting correlations well above the actual superconducting $T_c$, reminiscent of some of the pseudo-gap phenomena seen in underdoped cuprate superconductors \cite{pseudogap}.  This tendency is apparent in our results, where under many circumstances, the pairing scale is determined by interactions within a cluster, while $T_c$ is  proportional to $(t^\prime)^2$. As a consequence, pairing persists to a temperature $T_{pair} \sim (t^\prime)^0$, while the superfluid density, and hence the phase ordering temperature is parametrically smaller, $T_c \sim (t^\prime)^2$, as can be seen in Fig. 4.    
 
\section{The Inhomogeneous Hubbard Model}
While ideally we would like to consider a system in which any inhomogeneity is self-organized, in the present paper the inhomogeneity is introduced explicitly from the beginning. We therefore consider the Hubbard model on a square lattice 
\be \label{eq:ihm}
H=-\sum_{\left\langle \textbf{r},\textbf{r}'\right\rangle,\sigma}t_{\textbf{r},\textbf{r}'}\left(c^{\dagger}_{\textbf{r},\sigma}c_{\textbf{r}',\sigma}+ H.c.\right)+ U\sum_{\textbf{r}}n_{\textbf{r},\uparrow}n_{\textbf{r},\downarrow},
\ee
where 
$\left\langle \textbf{r},\textbf{r}'\right\rangle$ indicates nearest-neighbor sites, and $c^{\dagger}_{\textbf{r},\sigma}$ creates an electron on site $\textbf{r}$ with spin polarization $\sigma=\pm 1$. The term with $U>0$ represents the on-site repulsion between electrons and $n_{\textbf{r},\sigma}=c^\dagger_{\textbf{r},\sigma}c_{\textbf{r},\sigma}$.
The usual (homogeneous) limit of this model is obtained by taking, $t_{\textbf{r},\textbf{r}'}=t=1$, where the final equality defines our units of energy. In the inhomogeneous versions of this model we consider, the lattice is broken up into a set of periodically repeated disconnected clusters, with $t_{\textbf{r},\textbf{r}'}=1$ for nearest-neighbor sites within a single cluster, and $t_{\textbf{r},\textbf{r}'}=t^\prime \ll 1$ for  nearest-neighbor sites belonging to distinct clusters.   Even the inhomogeneous version of this model is particle-hole symmetric; we will discuss the case of a concentration $x$ of doped holes, but the same results apply for the same concentration of doped electrons.

To zeroth order in $t^\prime$, the Hamiltonian can be solved for arbitrary $U$ by diagonalizing 
it on a single cluster. We then use low order (near) degenerate perturbation theory (in powers of $t^\prime$) to derive an effective Hamiltonian, $H^{eff}$, which operates in the reduced Hilbert space spanned by the direct products of the low-energy 
eigenstates of the isolated cluster.  This is a standard procedure, precisely analogous to that used to derive the \textit{t-J} model from the large $U$ limit of the Hubbard model \cite{assa}.   
For all the clusters we consider here, the groundstate of the isolated undoped cluster (with one electron per site),   is a spin singlet with a finite spin gap $\Delta_{s}$.  

For small $x$, most clusters must still be in their groundstate, so $H^{eff}$ operates in a very much smaller Hilbert space than the starting space.  Moreover, defining the unique ground-state with one electron per site to be the vacuum state of $H^{eff}$, it is clear that it can typically be recast as the Hamiltonian of a dilute gas of excitations.  This is the key feature that makes the problem tractable in the stated limit of small $x$ and small $t^\prime$.  

To construct the low energy Hilbert space, we need to compute the spectrum of an isolated cluster with different numbers of doped holes. The eigenstates of each cluster can be identified by their symmetry related quantum numbers: the number of doped holes, $Q$ ($Q=0$ refers to the case of one electron per site) the total spin, $S$, and those related to the spatial symmetries. For the dimer, the states are odd or even under reflection. The isolated square has the same four-fold rotational symmetry, $C_4$, as the uniform lattice so the states can be labeled by spectroscopic labels ``$s$'' (even under rotation by $\pi/2$) ``$d$''  (odd under rotation by $\pi/2$) and ``$p_x\pm ip_y$'' (changes phase by $\pm \pi/2$ under rotation by $\pi/2$). In each charge sector, so long as there is a ``large'' (order 1) gap, the excited states can be safely eliminated from the low energy sector.  Where there is a level crossing within the isolated cluster, we need to be a bit more careful.

{\it Isolated dimer:}  For the isolated dimer with $Q=0$ or $Q=2$, there is a unique $S=0$, even parity ground state separated by a large gap from the first excited state. The $Q=1$ ground-state is a $S=1/2$ even parity doublet again with a large gap. 

{\it Isolated square:}  For the isolated square with $Q=2$, there is a unique $S=0$ ground state with $s$-wave symmetry separated by a large gap from the first excited state. For $Q=0$ and with $U=0$, there is a large (6-fold) ground-state degeneracy. This degeneracy is lifted \cite{sudip} at non-zero $U$, and the resulting ground-state is an $S=0$ singlet with $d$-wave symmetry.  However, for small enough $U$ the gap to the excited states is small -- the splitting between the lowest lying singlet and triplet state (the ``spin-gap'') is ${\cal O}(U^2)$ for small $U$.  In the present paper, when dealing with the checkerboard lattice, we will assume $U \gg \sqrt{t^\prime}$ so that the gap can be treated as ``big'', but the small $U$ limit is probably worth revisiting in the future. As pointed out by Trugman and Scalapino \cite{trugman}, an important consequence of the distinct spatial symmetries of the $Q=0$ and $Q=2$ ground states is that the pair creation operator that connects these two-states has $d$-wave symmetry.  
The $Q=1$ spectrum of the isolated square is a bit more complex:  For $U < U_T= 18.6$, the ground-state is a spin and orbital doublet, with $S=1/2$ and $p_x \pm ip_y$ symmetry. This has the consequence that the quasiparticles carry an orbital ``flavor'' index, $\lambda=\pm 1$, in addition to the usual spin polarization index, $\sigma = \pm 1/2$.  For $U > U_T$, the ground state, in accordance with Nagaoka's theorem \cite{Nagaoka}, is a $S=3/2$ $s$-wave state.  Except in the vicinity of $U=U_T$ (where the gap is ${\cal O}(|U-U_T|)$, the gap to excited states is again large.

\begin{figure}[t]
\subfigure[]{\includegraphics[width=6.1cm]{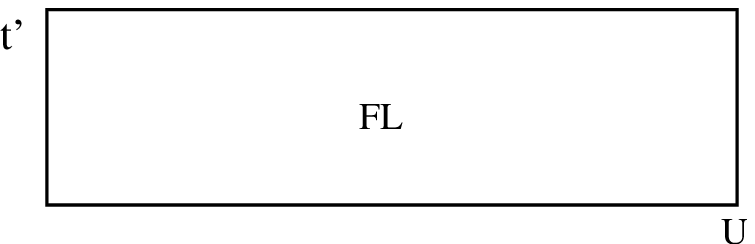}}
\subfigure[]{\includegraphics[width=6.5cm]{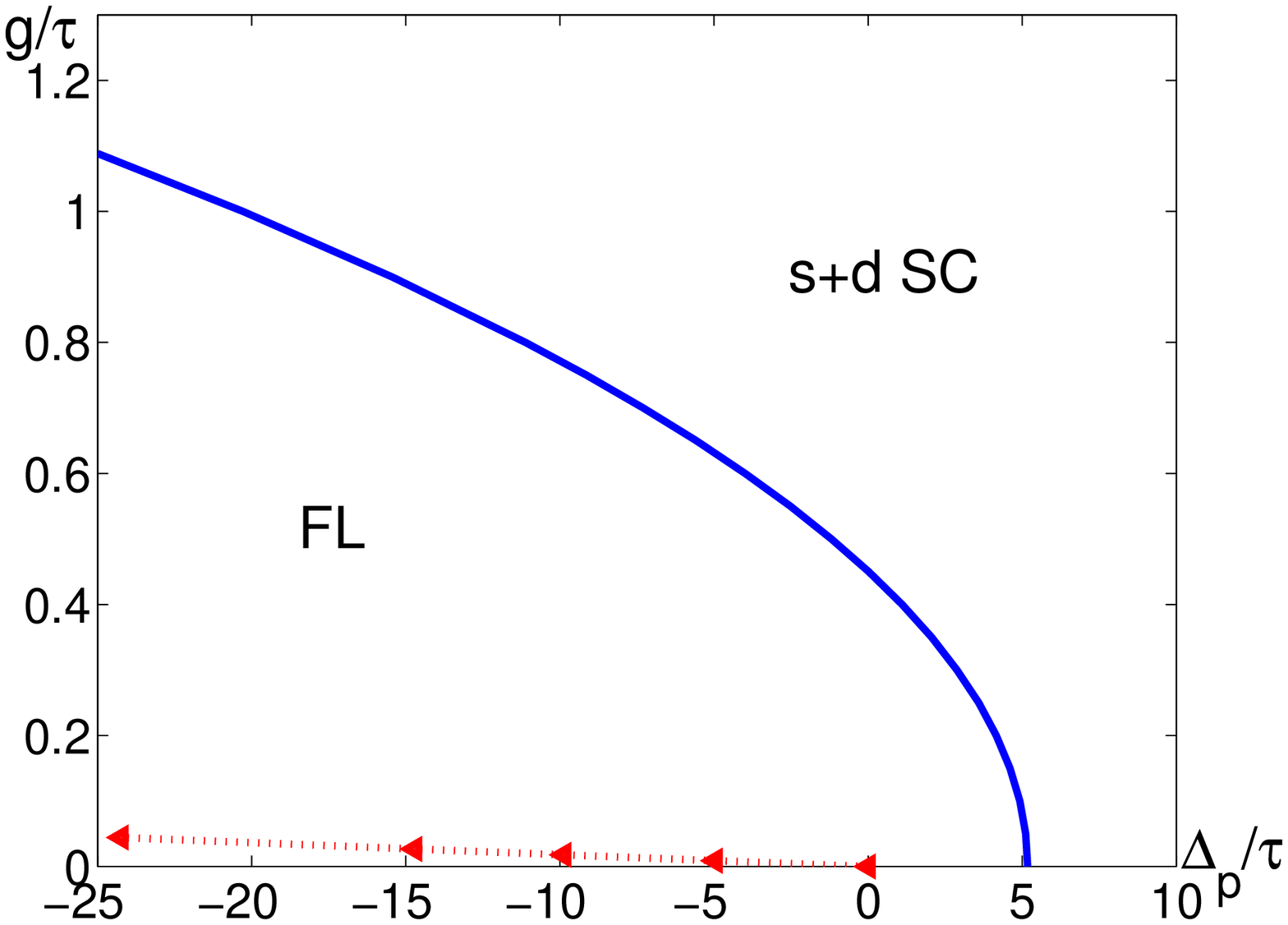}}
\caption{{\footnotesize (a) Zero-temperature phase diagram of the dimerized Hubbard model for small $x$, and (b) phase diagram of $H^{eff}$ on the dimerized lattice as functions of $\Delta_{p}/\tau$ and $g/\tau$. Two phases, the Fermi liquid phase (FL) and the singlete superconducting phase (s+d SC), can be obtained in the effective thoery. However, as shown by the dotted curve with arrows in (b), the trajectory of the Hubbard model with fixed $t'/t$ (=0.005) as a function of increasing $U$ only passes through FL. As a result, only FL can be seen in (a).}}
\end{figure}

Other than the stated level crossings, the precise dependence of the energies of the various states is not important for present purposes. The energy of the $Q=0$ groundstate can be absorbed into an overall constant contribution to the effective Hamiltonian, $E_0$, and the energy of the $Q=1$ state into a redefinition of the chemical potential. There is one important combination of energies, 
\be \label{eq:pbe}
\Delta_{p}=2E(1)-E(2)-E(0),
\ee
where $E(Q)$ is the ground-state energy for given $Q$. This has the interpretation of the pair binding energy: a positive $\Delta_{p}$ signifies an effective attraction between doped electrons or holes in the sense that for two doped holes, it is energetically preferable to place both on one cluster than to place one on each of two clusters. For the isolated square, $\Delta_p$ is positive (pair binding) for $U<U_c\approx 4.58$, and negative for $U>U_c$. For the isolated Hubbard dimer, $\Delta_p \le 0$ for all $U$, and indeed it vanishes (linearly) only at $U=0$.  This is the  reason the dimerized Hubbard model does not superconduct.

\section{The Effective Boson-Fermion Model}
It is now straightforward to obtain the effective Hamiltonian on the {\it cluster lattice} to first order in $t^\prime$ taking the unique ground-state of the undoped system as the vacuum state (See appendix for the derivation):
\ba \label{eq:eff1}
H^{eff} &=& E_0+\sum_{j}\left(-\Delta_{p}n_{bj} -\mu[2n_{bj}+n_{fj}]
\right)\nonumber \\
 &-&\sum_{\left\langle ij\right\rangle}\sum_{\sigma,\lambda\lambda^\prime}
\left(\tau_{ij,\lambda\lambda^\prime}
a^{\dagger}_{i,\sigma,\lambda}a_{j,\sigma,\lambda^\prime}+H.c.\right)
\nonumber \\
&+& \sum_{\left\langle ij\right\rangle,\lambda\lambda^\prime}\left(g_{\lambda\lambda^\prime}\phi_{ij}b^{\dagger}_{i}\left[a_{i,\uparrow,\lambda}a_{j,\downarrow,\lambda^\prime}-(\uparrow\leftrightarrow\downarrow)\right]+H.c.\right) \nonumber \\
&+& U^{\infty}\sum_{i}\left(n_{fi}+n_{bi}\right)\left(n_{fi}+n_{bi}-1\right)
\ea 
where $a^{\dagger}_{i,\sigma,\lambda}$ creates an one-hole fermion on the $i$th cluster with spin polarization $\sigma$ and (for the checkerboard case) flavor index $\lambda$ and $b^{\dagger}_{i}$ creates a hole-pair boson on the  $i$th cluster; $n_{bj}=b^\dagger_jb_j$ and $n_{fj}=\sum_{\sigma,\lambda}a^\dagger_{j,\sigma,\lambda}a_{j,\sigma,\lambda}$ are, respectively, the boson and fermion densities on cluster $j$.  The coupling constants, $\tau$, $g$ (both proportional to $t^\prime$), and our old friend $\Delta_{p}$ represent the effective hopping of one-hole fermions, the fermion-boson Andreev coupling, and the energy difference between one  boson and two fermions respectively. $\langle ij \rangle$ represents a pair of nearest neighbor clusters, and $\phi_{ij}$ is a geometric factor discussed below.  The effective Hamiltonian operates in a constrained Hilbert space where $n_{bj}+n_{fj} = 0$ or $1$, but equivalently, it can operate in an unconstrained Hilbert space with the constraint imposed dynamically by taking the limit $U^\infty\to\infty$.  
The chemical potential is, of course, an implicit function of the doping concentration $x$ of the original model, obtained by inverting the relation $\sum_{i}(n_{fi}+2n_{bi})=Mx$, where $M$ denotes the number of lattice sites in the original Hubbard model.

This effective Hamiltonian is not only built on a highly reduced Hilbert space, but also has common structure for both models of our current study. It has a fermion band (or multi-bands with index $\lambda$), a boson band (with infinite bare effective mass) and the interactions between them which converts a pair of fermions to a boson and vice versa. The effective theory is analogous in form to the so-called ``Boson-Fermion'' model which has been studied by several people \cite{micnas}. However, because in the present case, this model is derived as the low-energy effective field theory from the inhomogeneous Hubbard model, we are lead to study it in particular limits (especially $x\ll 1$) that were not the focus of previous studies.

For the dimerized model, the dimers explicitly break the $C_4$ symmetry of the underlying lattice, and hence $H^{eff}$ only has $C_2$ symmetry.  Explicit evalution of the first order perturbation theory leads to the expressions $\Delta_p = -2t(1+\text{tan}\theta)$, $g=(\text{cos}\theta+\text{sin}\theta)t^\prime$, and for nearest neighbor dimers in the same column of dimers,  $\tau_{ij}=(1-\text{sin}2\theta)t^\prime/2$ and $\phi_{ij}=1$, while for dimers in neighboring columns, $\tau_{ij}=(1-\text{sin}2\theta)t^\prime/4$ and $\phi_{ij}=1/2$ where $\text{tan}2\theta=-4t/U$. 

\begin{figure}[t]
\subfigure[]{\includegraphics[width=6.4cm]{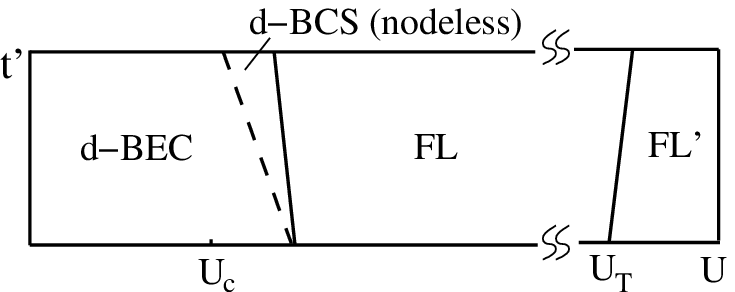}}
\subfigure[]{\includegraphics[width=6.8cm]{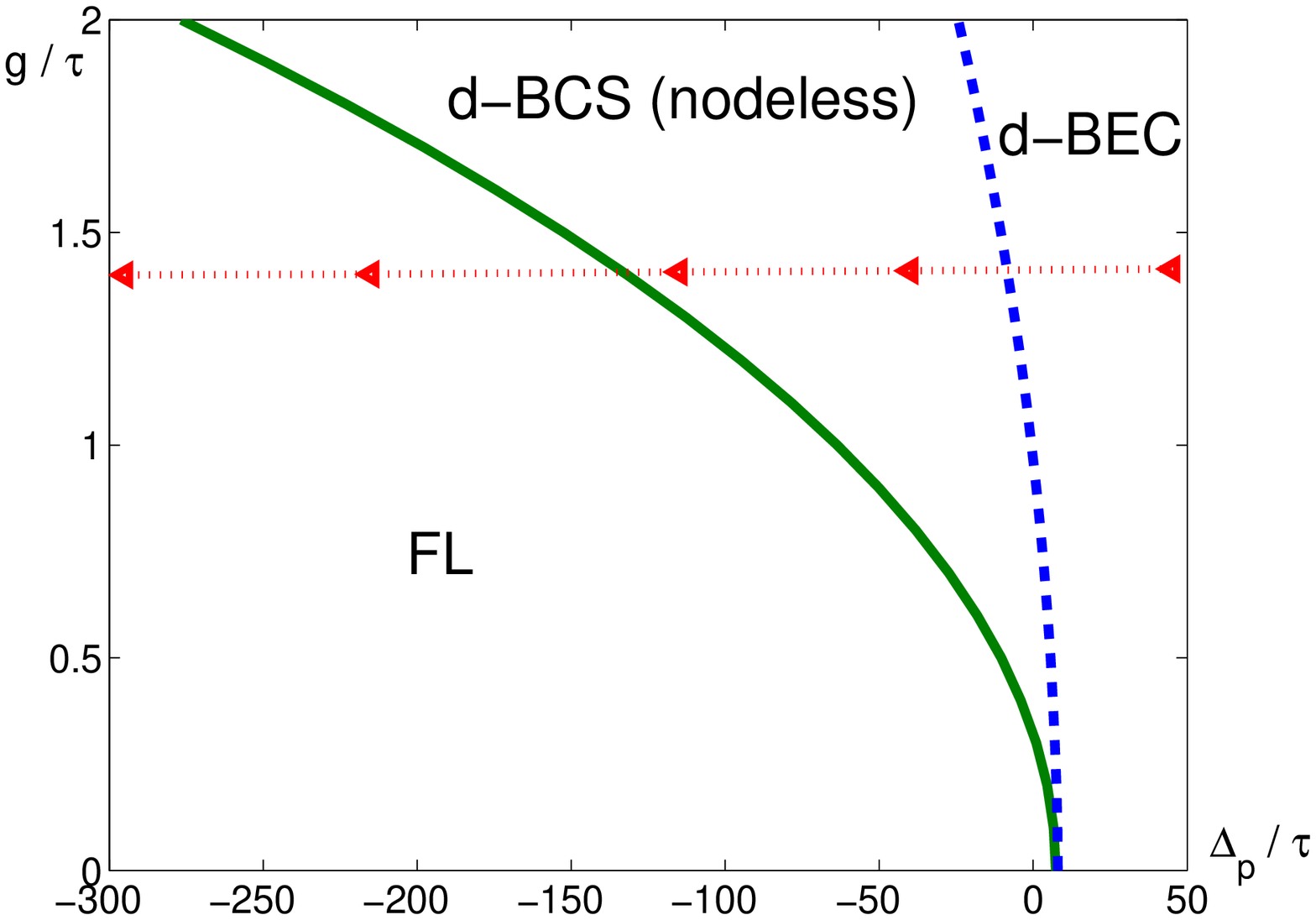}}
\caption{{\footnotesize (a) Zero-temperature phase diagram of the checkerboard Hubbard model for small $x$(=0.025) and (b) of $H^{eff}$ on the checkerboard lattice. The two superconducting phases
both have $d$-wave symmetry and \textit{no} nodal quasi-particles, but one is  BCS-like ($d$-BCS) and the other is a Bose condensate of preformed two-particle bound-states ($d$-BEC).  
The dashed lines indicate a crossover rather than a sharp phase transition. Notice that the boundary (not drawn here) for the presence of a two-particle bound state in Eq.(\ref{eq:eff1}) is close to the crossover curve.  The two Fermi liquid phases (FL and FL') are distinguished by the quantum numbers of the quasi-particles. The dotted curve with arrows in (b) represents the effective trajectory corresponding to the Hubbard model with fixed $t'/t$ (=$0.005$) as a function of increasing $U$.}}
\end{figure}

The checkerboard model preserves the $C_4$ symmetry of the underlying lattice.  The eigenvalue problem on the isolated square is more complex, so the expressions for the effective couplings are somewhat complicated. Some details are shown in the appendix.  To give representative values, we give expressions for $U$ in the neighborhood of $U=U_c$ where $\tau_{ij,\lambda\lambda^\prime}= i\tau\phi_{ij} \epsilon_{\lambda\lambda^\prime}$ with $\tau  =[\alpha_{0}+\alpha_{1}(U-U_c)+\ldots ]t^\prime$  for all pairs of nearest-neighbors ($\alpha_{0}=0.2217$, $\alpha_{1}=-0.0081$), and $g_{\lambda\lambda^\prime}=g[1-\delta_{\lambda,\lambda^\prime}]$
where $g=[\beta_{0}+\beta_{1}(U-U_c)+\ldots]t^\prime$  and $\beta_{0}=0.3123$, $\beta_{1}=-0.0111$, and as a consequence of  the $d$-wave symmetry of the pair creation operator $\phi_{ij}=1(-1)$ for neighboring squares along the x-axis (y-axis).  (Here $\epsilon_{\lambda\lambda^\prime}$ is the Levi-Civita tensor.)

\section{$T=0$ Phase Diagrams of the dimerized and the checkerboard Hubbard models }
Let us now consider the phase diagram of $H_{eff}$.  As mentioned above, for small $x$, the density of excitations is small, so that two particle collisions are rare (order $x^2$) and multiparticle processes even rarer.  Thus, when $|\Delta_p| \gg t^\prime$, the phases are pretty obvious:  

\noindent{\bf 1)} For $\Delta_p < 0$, the bosons appear only as virtual states, leading to a weak induced attraction between fermions of order $V^{ind} \sim g^2/|\Delta_p|$. However, because of the hard-core repulsion between fermions, the net interaction is repulsive, and hence there is a Fermi liquid phase. (There could be some form of Kohn-Luttinger instability \cite{Chubukov} at exponentially low temperature, but we will not worry about this.) For the dimerized model, the Fermi surface is a small ellipse closed about $\vec k=( \pi/2a,\pi/a )$ where $a$ is the lattice constant of the original square lattice (FL in Fig. 2(a)).  In the checkerboard model with $U_c < U < U_T$, there are two small Fermi surface circles (due to the flavor degeneracy);  one is closed about $\vec k=(0,\pi/2a )$ and the other about $\vec k=(\pi/2a,0)$ (FL in Fig. 3(a)).   
For $U > U_T$, the system forms a Fermi liquid of spin 3/2 fermions (FL' in Fig. 3(a)).
There is a (presumably first order) 
transition between these two sorts of Fermi liquid which occurs at $U =U_T +{\cal O}(t^\prime)$.

\noindent{\bf 2)} For $\Delta_p >0$, it is the charge $Q=1$ quasiparticles that are virtual excitations.  Integrating them out generates an effective nearest-neighbor hopping matrix $\tau_b^{eff}\sim g^2/\Delta_p$, and a nearest-neighbor boson-boson repulsion of the same order \cite{steve}. Thus, in this limit, the system has a singlet superconducting groundstate.  In the dimerized model
there is no other symmetry classification of the superconducting state possible (it is an admixture of $d$-wave and $s$-wave states), but for the checkerboard model, the superconducting state inherits the $d$-wave symmetry of the constituent bosons. This $d$-wave symmetry could be observed in any of the phase sensitive measurements that have been discussed in the context of the cuprates themselves.  However, it is important to realize that, for both the dimerized and checkerboard model, there are no gapless spin 1/2 excitations, and hence no nodal quasiparticles.  In the language of BCS theory, one can think of the $d$-wave state in the checkerboard lattice as being in a strong-coupling limit in which the chemical potential has passed below the band-bottom.

There are several features of the thermal evolution of the system in this limit that warrant mention. The first is that, since it is a system of preformed bosons, $T_c$ is determined by the zero temperature superfluid density, $T_c\sim \tau_b^{eff} x\sim (t^\prime)^2\Delta_p^{-1} x$.  Of course, since the model is two dimensional, this also means that there should be a 
vortex gas regime above $T_c$ with a Kosterlitz-Thouless transition to a phase with quasi-long-range superconducting order. In addition, there can be two other pseudo-gap scales apparent up to temperatures which are parametrically larger than $T_c$:  Firstly, $T_p \sim \Delta_p/|\ln(2x)|$ is the characteristic temperature at which pairs thermally dissociate.  Above this, there is a temperature $T^\star \sim {\cal O}(t'^{0})$, at which excitations beyond those in the effective model become significant. Our finite temperature results discussed above are summarized in a schematic diagram, Fig. 4.

\begin{figure}[t]
\includegraphics[width=6.8cm]{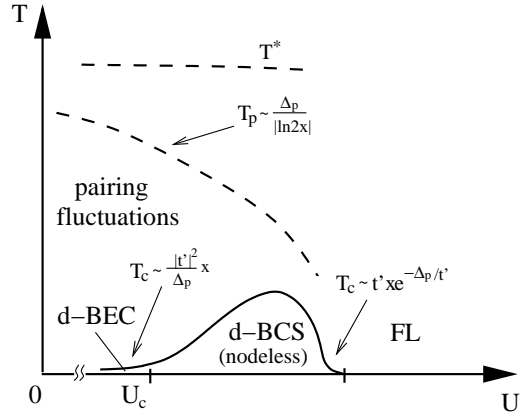}
\caption{{\footnotesize Schematic finite temperature phase diagram of the checkerboard model. Notice that this figure is not drawn to scale and the x-axis has been offset from zero. As in Fig. 3(a), 
there is a crossover within the superconducting phase from a $d$-BEC to a $d$-BCS limit.  For large enough $U$ there is a FL phase.  
Above $T^{*}\sim\mathcal{O}(t'^{0})$, indicated by the top dashed curve, the effective theory breaks down. The lower dashed curve, $T_p$, signifies the crossover between 
a normal Bose liquid of preformed pairs for $T < T_p$ and
a high temperature state with little electron pairing for $T > T_p$.  
The range $T_c < T < T_p$ is a pseudo-gap regime. 
This can be a broad regime since the energy scale of coherence between preformed pairs is $\mathcal{O}(t'^2)$ as indicated by the solid curve.}}
\end{figure}

\noindent{\bf 3)} The only part of the phase diagram which is at all subtle is where $\Delta_p \sim t^\prime$, where both bosonic and fermionic excitations participate in the low energy physics.  However, this is also in some ways the most interesting regime.  In the first place, approaching this regime from $\Delta_p >0$, we see that the superconducting $T_c \propto 1/\Delta_p$ reaches its maximum in this regime.  Moreover, it is only in this regime that we could possibly have a superconducting state with gapless nodal quasi-particles (although none occur in this model, unfortunately).  

The feature that makes this regime tractable is that the transition from the Fermi liquid to the superconducting phase is BCS-like.  This is because, for small $x$, it is only the pairwise interactions between quasi-particles that are relevant.  The effective interaction is the sum of the
induced attraction, $V^{ind}$, and the hard-core repulsion between quasiparticles, which, treated in the T-matrix approximation, gives rise to an effective repulsion $V_0 \sim t^\prime$. Where the sum, $V_T$ of these two terms switches from being a net repulsive to a net attractive interaction, it passes through a transition point at which $V_T=0$.  Even though the Fermi energy is only of order $t^\prime$, arbitrarily near the transition point, $|V_T| \ll E_F$, justifying the use of BCS mean-field theory. For $H^{eff}$ on the dimerized lattice, since the superconducting state has a substantial $s$-wave component, we generally expect the superconducting state near the critical value of $\Delta_p$ to be nodeless, and thus adiabatically connected to the superconducting phase at large $\Delta_p$.  On the other hand, for the checkerboard lattice, the superconducting state has $d$-wave symmetry, but a feature of the Fermi surfaces enclosing $(\pi/2a,0)$ and $(0,\pi/2a)$ of the BZ is that the line of gap zeros does not intersect them, and consequently there are no gapless quasi-particles in this phase. As $\Delta_p$ gets larger, the superconducting state smoothly evolves from a BCS-like region, where quasi-particle fluctuations are dominant ($d$-BCS), to a regime of pre-formed pairs, where phase fluctuations of the order parameter dominate the physics ($d$-BEC). This is quite distinct from a $d$-BCS phase with nodal quasi-particles.

To obtain the explicit phase diagram for $H^{eff}$
presented in Fig. 2 and 3, we have made use of a mean-field approximation which, none-the-less, reproduces all the qualitative features required by the above qualitative analysis.  Specifically, we perform a standard mean-field decomposition of $H^{eff}$, and solve self-consistently for the anomalous expectation values $\langle b_j^\dagger\rangle$ and $\langle a_{j,\uparrow}^\dagger a_{i,\downarrow}^\dagger\rangle$. To avoid repetition, we would like to refer the readers to references\cite{micnas, Domanski}.
In this analysis, the hard-core constraint for the bosons is treated exactly, but the onsite repulsion involving the fermions is treated as an effective repulsion,  $V_0\sim t^\prime$;  predictably, the results do not depend substantially on $V_0$.

On the dimer lattice, the phase diagram of $H^{eff}$ as a function of $\Delta_{p}/\tau$ and $g/\tau$, shown in Fig. 2(b), exhibits a Fermi liquid and a nodeless $d+s$ superconducting phase.  If we consider only the allowable region of parameters that can be derived from the dimerized Hubbard model, these always lie within the Fermi liquid phase as indicated by the dotted curve in Fig. 2(b). The arrow on the curve represents the direction of change as $U$ increases from 0 to a large value. If we had instead considered the dimerized $t$-$J$ model, then for $J/t > 2-22.9(t'/t)$
the system is superconducting.

The phase diagram of 
$H^{eff}$ on the checkerboard lattice is shown in Fig. 3(b). 
The dotted curve in Fig. 3(b) is the trajectory through the phase diagram derived from the checkerboard Hubbard model with fixed   $t'=0.005$, where the arrow indicates the direction of change as $U$ increases from 0 to a large value not too close to $U_T$. 
It is worth emphasizing that, in this case, superconductivity with $d$-wave symmetry is obtained from purely repulsive interaction and can survive in the region with slightly negative $\Delta_{p}$ ($\sim \mathcal{O}(-t^{\prime})$).

\section{Perspective}
In the present study,  the translational symmetry of the lattice is explicitly broken and the inhomogeneity is taken to be large, $t'\ll t$. While various sorts of inhomogeneities have been found to be widespread in the cuprates, it is currently unclear if they are relevant for  the mechanism of the high temperature superconductivity. In part, this is a theoretical issue that hinges on the still unsettled issue of what is the superconducting $T_c$ (if any) of the uniform Hubbard model.  Clearly, when there is a superconducting groundstate in the highly inhomogeneous limit ($t^\prime/t \ll 1$), $T_c$ is generally an increasing function of $t^\prime$.  If $T_c$ is small or 0 in the homogeneous Hubbard model, there must be an intermediate value of $t^\prime$ (``optimal inhomogeneity'') at which $T_c$ is maximized, as suggested in Ref. \onlinecite{Arrigoni,ivar,fradkinreview}.  If this is the case, it is suggestive that the self-organized inhomogeneities found in the cuprates may be important for the mechanism of superconductivity.
\begin{acknowledgments} 
We thank Tim Goodman who alerted us to the possibility of a sign error in $H^{eff}$ with an insightful symmetry argument in our previous version of the manuscript. The work was supported, in part, by NSF Grant DMR-04-21960  at Stanford and DOE Grant DE-FG03-00ER45798 at UCLA.
\end{acknowledgments} 

\appendix
\section{A derivation of the effective theory for the checkerboard Hubbard model}
Taking the checkerboard Hubbard model as an example, here we would like to show how the effective Hamiltonian can be derived by using (degenerate) perturbation theory. To begin with, it is necessary to determine the low-energy, reduced Hilbert space.
All eigenstates of a 4-site Hubbard model can be calculated analytically by taking advantage of the symmetries of the model. Although conceptually simple, there are totally 256 eigenstates, making the actual calculations involved.
Fortunately, the results have been published by R. Schumann\cite{schumann} and explicit eigenstates can be found on his website. For $U\sim U_c$, the chosen lowest energy states in each charge $Q$ sector are summerized (using Schumann's notation) in Table I. 

\begin{table}[b]
\begin{center}
\caption{The eigenstates and eigenvalues of a 4-site Hubbard model in the low-energy Hilbert space. The corresponding charge secter (Q), total spin (S), spin $z$-component ($S_{z}$), symmetry under $\pi/2$ rotation, and eigenstate number used in Schumann's paper are also given.
($\text{cos}\alpha=[(36t^2 U-U^3)/27]/[(48t^2 +U^2)/9]^{3/2}$, $\text{cos}\beta=4t^2 U/[(16t^2+U^2)/3]^{3/2}$)}
\begin{tabular}{|c|c|c|c|c|c|} \hline
\hline
Q  & E(Q) & S & $S_{z}$ & symmetry & eigenstate \\ \hline 
& & & & &  \\
0    & $\frac{\sqrt{3}U-2\sqrt{16t^2 +U^2}\text{cos}(\frac{\beta}{3})}{\sqrt{3}}$ & 0  & 0  & d-wave & $\Psi_{111}$ \\ 
& & & & &  \\ \hline
 & & & & & $\Psi_{46}$,$\Psi_{70}$ \\
1    & $\frac{U-\sqrt{32t^2 +U^2 +4\sqrt{64t^4 +3t^2 U^2}}}{2}$ & $\frac{1}{2}$ & $\pm\frac{1}{2}$  &  $p_{x}\pm ip_{y}$ & \\  
& & & & & $\Psi_{50}$,$\Psi_{74}$ \\ \hline
& & & & &  \\
2    & $\frac{U-2\sqrt{48t^2 +U^2}\text{cos}(\frac{\alpha}{3})}{3}$ & 0 & 0  &  s-wave  &  $\Psi_{22}$ \\ 
& & & & &  \\ \hline
\end{tabular}
\end{center}
\end{table}

For $t^\prime=0$ and $1\gg x >0$, the groundstate of the model is highly degenerate since the doped holes can be distributed among the decoupled squares in many ways. To an the effective model in powers of $t^\prime$, we employ standard degenerate perturbation theory 
Write Eq.(\ref{eq:ihm}) as $H=H_{0}+H^\prime$ ($H^\prime$ represents the $t'$ term) and let $\mathcal{P}$ be the projection operator onto the subspace spanned by the direct product of the states shown in Table I.  Then 
\begin{equation} 
\label{eq:dpt}
H^{eff}=\mathcal{P}H\mathcal{P}+\mathcal{P}H^\prime [ 1-\mathcal{P}]\frac{1}{E_{0}-H_{0}}[1-\mathcal{P}]H^\prime \mathcal{P} + \ldots.
\end{equation}

For our present study, we keep only terms to first order in $t'$. 
Specifically, labelling the states in accord with the notation of Ref.\onlinecite{schumann}, 
the various matrix elements of $H^\prime$ between the unperturbed groundstates can be expressed as follows: the only non-vanishing matrix elements are 1) elements related to the effective hopping amplitude $\tau$,
\begin{eqnarray}
i\tau &=& \left\langle \Psi_{111}(j')\right|\left\langle \Psi_{46}(j)\right|H'\left|\Psi_{111}(j)\right\rangle\left|\Psi_{50}(j')\right\rangle \nonumber \\
 &=& -\left\langle \Psi_{111}(j')\right|\left\langle \Psi_{50}(j)\right|H'\left|\Psi_{111}(j)\right\rangle\left|\Psi_{46}(j')\right\rangle \nonumber \\
&=& \left\langle \Psi_{111}(j')\right|\left\langle \Psi_{70}(j)\right|H'\left|\Psi_{111}(j)\right\rangle\left|\Psi_{74}(j')\right\rangle \nonumber \\
&=&- \left\langle \Psi_{111}(j')\right|\left\langle \Psi_{74}(j)\right|H'\left|\Psi_{111}(j)\right\rangle\left|\Psi_{70}(j')\right\rangle, \nonumber \\
\end{eqnarray} 
with $j,j'$ representing nearest-neighbor unit cells;
2) elements related to the boson-fermion coupling $g$,
\begin{eqnarray}
g &=& \left\langle \Psi_{111}(j')\right|\left\langle \Psi_{22}(j)\right|H'\left|\Psi_{46}(j)\right\rangle\left|\Psi_{74}(j')\right\rangle \nonumber
\\
&=& \left\langle \Psi_{22}(j')\right|\left\langle \Psi_{111}(j)\right|H'\left|\Psi_{46}(j)\right\rangle\left|\Psi_{74}(j')\right\rangle \nonumber \\
&=& \left\langle \Psi_{111}(j')\right|\left\langle \Psi_{22}(j)\right|H'\left|\Psi_{50}(j)\right\rangle\left|\Psi_{70}(j')\right\rangle \nonumber \\
&=& \left\langle \Psi_{22}(j')\right|\left\langle \Psi_{111}(j)\right|H'\left|\Psi_{50}(j)\right\rangle\left|\Psi_{70}(j')\right\rangle \nonumber \\
&=& -\left\langle \Psi_{111}(j')\right|\left\langle \Psi_{22}(j)\right|H'\left|\Psi_{74}(j)\right\rangle\left|\Psi_{46}(j')\right\rangle \nonumber \\
&=& -\left\langle \Psi_{22}(j')\right|\left\langle \Psi_{111}(j)\right|H'\left|\Psi_{74}(j)\right\rangle\left|\Psi_{46}(j')\right\rangle \nonumber \\
&=& -\left\langle \Psi_{111}(j')\right|\left\langle \Psi_{22}(j)\right|H'\left|\Psi_{70}(j)\right\rangle\left|\Psi_{50}(j')\right\rangle \nonumber \\
&=&- \left\langle \Psi_{22}(j')\right|\left\langle \Psi_{111}(j)\right|H'\left|\Psi_{70}(j)\right\rangle\left|\Psi_{50}(j')\right\rangle, \nonumber \\
\end{eqnarray} 
Explicit expressions for these matrix elements as a function of $U$ can be obtained using the explicit wavefunctions given in Ref.\onlinecite{schumann} (See Fig. 5).
When expressed in second quantized form, the resulting effective Hamiltonian is given in Eq. (\ref{eq:eff1}), above.  

\begin{figure}[b]
\includegraphics[width=6.8cm]{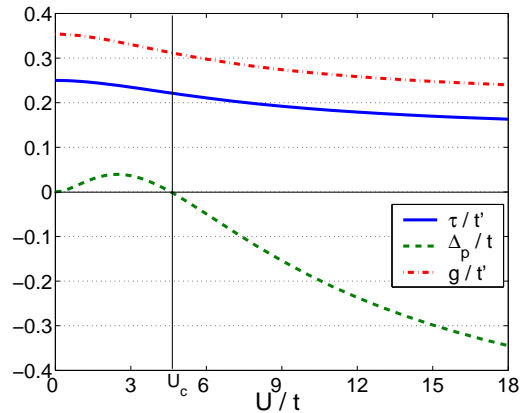}
\caption{{\footnotesize The effective parameters, $\tau/t'$, $g/t'$, and $\Delta_p /t$ as a function of $U/t$.
}}
\end{figure}


\begin{thebibliography}{20}
\bibitem{steve}
Strictly speaking, other interactions of order $(t^\prime)^2$ should be included in this analysis except in the immediate vicinity of a critical value of $U$ at which $\Delta_p$ vanishes, but these are qualitatively similar to the ${\cal O}(t^\prime)^2/\Delta_p$ interactions discussed here, and their inclusion has no effect on our qualitative conclusions.
\end{thebibliography}
\end{document}